\documentclass[10pt]{llncs}

\usepackage{graphicx}
\usepackage{paralist}
\usepackage{subfig}
\usepackage{todonotes}
\usepackage[urlcolor=blue]{hyperref}
\usepackage[nodayofweek,level]{datetime}
\usepackage{cleveref}
\crefformat{footnote}{#2\footnotemark[#1]#3}

\let\llncssubparagraph\subparagraph
\let\subparagraph\paragraph
\usepackage[compact]{titlesec}
\let\subparagraph\llncssubparagraph

\begin{document}


\title{Lessons Learnt from a 2FA roll out within a higher education organisation}


\author{
	Abideen Tetlay\inst{1} \and
	Helen Treharne\inst{1} \and
	Tom Ascroft\inst{1} \and
	Sotiris Moschoyiannis\inst{1} 
}
\institute{
 University of Surrey, UK,\\
	\email{\{abideen.tetlay,h.treharne,t.ascroft,s.moschoyiannis\}@surrey.ac.uk}
}

\pagestyle{plain}

\maketitle

\begin{abstract}
Rolling out a new security mechanism in an organisation requires planning, good communication, adoption from users, iterations of reflection on the challenges experienced and how they were overcome.  Our case study elicited users' perceptions to reflect on  the adoption and usage of the two factor authentication (2FA) mechanism being rolled out within our higher education organisation. This was achieved using a mixed method research approach.  Our qualitative analysis, using content and thematic coding, revealed that initially SMS was the most popular 'second factor' and the main usability issue with 2FA was the getting the authenticator app to work; this result was unexpected by the IT team and led to a change in how the technology was subsequently rolled out to make the authenticator app the default primary second factor. Several lessons were learnt about the information users needed; this included how to use the technology in different scenarios and also a wider appreciation of why the technology was beneficial to a user and the organisation. 
The case study also highlighted a positive impact on the security posture of the organisation which was measure using IT service request metrics.

\end{abstract}

\section{Introduction}For organisations, one of the drivers causing adoption of stronger authentication is the increasing cost of data breaches, with a global average cost per breach of \$3.92m~\cite{IBM:Breach:2019}, a rise in cost by 12\% over 6 years, and with one survey reporting that 94\% of organisations cite that data breaches in the previous 12 months have influenced their security policies~\cite{Thales:Access}. With malicious attacks causing 51\% of data breaches in 2019~\cite{IBM:Breach:2019}, organisations need stronger protection. In general, 58\% of organisations believe that two-factor authentication (2FA) is the most likely access control tool that will be used to protect their systems~\cite{Thales:Access}, while 49\% believe it is single sign-on.  The UK Government guidance has clearly set out several guidance documents on how to implement multi-factor authentication (MFA) as an organisation~\cite{NCSC:2FAimp} and also how to set up 2FA~\cite{NCSC:2FA}. It is recommended that 2FA is used for accounts such as your bank account, patient records, etc., where if the passwords are stolen would cause most harm. When using 2FA users will be asked to provide a second factor for example through the use of authenticator applications, text messages (SMS) or other second factors.

The University of Surrey rolled out 2FA to strengthen its resilience against data breaches and to improve compliance on how data was being accessed from university systems between 2019 and 2020 and is continuing to do so. In this case study we evaluate the adoption of 2FA within our University from two perspectives. Firstly, we explore the users' perceptions, adoption and usage of 2FA. Our findings identified the need for improved cyber awareness of users in order to appreciate some of the inconveniences related to 2FA. Secondly, we report on the insights and lessons learnt on the user support and guidance needed from an IT deployment perspective. 
It is the combination of users' perspective of using the technology and the perspective of the support users needed from IT which is the interesting HCI aspect of this case study since there are trade-offs between what users want to experience and what the IT deployment allows.

 Our user findings echo many of those already found in the academic literature regarding the ease of use, the difficulties in setting up the second factor, and users' concerns when things go wrong. 
 We identified the following research questions for our case study:
 
\begin{description}
\item[Question 1:]
What is the most popular ‘second factor’ for 2FA within the University?
\item[Question 2:]
 What is the least popular ‘second factor' for 2FA within the University?
\item[Question 3:]
What are the main usability issues with 2FA within the University?
\item[Question 4:]
What do 2FA users value the most from the following {\it Usability, Security, Privacy, Trust} in the context of 2FA?  
\item[Question 5:]
What are the lessons learnt from an IT deployment perspective and impact on the University from a user support perspective?
\end{description}

The remainder of this paper is structured as follows. Section~\ref{sec:relatedwork} outlines the related work. Section~\ref{sec:design} describes the research design for our case study including limitations. Section~\ref{sec:timeline} provides an overview of the 2FA deployment at the University of Surrey. Section~\ref{sec:findings} presents the findings from our user evaluations and Section~\ref{sec:ITfindings} provides the lessons learnt from an IT perspective to improve the deployment of the 2FA security mechanism. Section ~\ref{sec:discussion} provides an overall discussion of the insights and Section~\ref{sec:conc} provides concluding remarks. 

\section{Related Work}
\label{sec:relatedwork}
This section provides an overview of several 2FA studies that have been conducted in the academic literature and by industry.  An industry survey,~\cite{Thumb:Authentication}, conducted by an industry group: ThumbSignIN, One World Identity and Identify and Access Managment provided Gluu, observed than 60\% of the companies surveyed are already using strong authentication and 29\% are looking to implement or expand their use of 2FA and 83\% recognise the need to do this for better security posture. The report notes that the main industry driver is the need to be compliant when services related to the EU's Revised Payment Services Directive are being offered mandate the use of 2FA. A yearly report by LastPass in 2019 noted that 57\% of the 47,000 businesses surveyed are using MFA, a 12\% increase on the previous year~\cite{LastPass:Security}. Of the employees of those businesses using MFA 95\% used a software-based multifactor authenticator as the second factor and only 4\% a hardware-based solution. It was encouraging to note that the education sector's use of MFA (with 33\% adoption) was the second business sector listed for its use of MFA after technology/software but ahead of banking/finance. The report also noted that adoption of 2FA was more commonplace in larger industries.

The majority of the academic case studies have been in the U.S. by Colnago et al. ~\cite{Colnago:CMU}, Weidman and Grossklags~\cite{weidman:asac2017} and Abbott and Patil ~\cite{abbott:chi2020}, and focused on themes of transitioning to 2FA, the trade-off between user experience (UX) and security, and the adoption of 2FA with bring your own device (BYOD).

Colnago et al. distributed surveys to staff and students at Carnegie Mellon University (CMU) prior to and after roll out of 2FA, whereas in our user study it was after roll out.  For staff at CMU, 2FA was mandatory yet by the implementation deadline, only 75\% had enrolled. Enrolment was optional for students unless they used the University payroll system, and this may account for the relatively high 50\% enrolling by the deadline. Their findings highlighted that, prior to adoption of any form of 2FA, users were concerned about its usefulness, usability and inconvenience, especially the perceived increase in time spent logging in. However, after adoption, users tended to find that it was easy to use. The paper highlighted the IT support cost leading up to implementation, where 24\% of IT tickets were 2FA related, whereas we identified the benefit following roll out.

Weidman and Grossklags examined user attitudes of staff when adopting 2FA at Pennsylvania State University comparing the use of a hardware token with a DuoMobile authenticator app and the emphasis of the study was on utilising employees own devices. Their findings noted that the ease of use was greater for the hardware token than the app but that the app was more compatible with the workflow of participants. Their findings noted users' concerns over using their own devices, a concern echoed in our findings.

Abbott and Patil focused on analysing online surveys for both students and staff in a U.S university and system logs during a multi-phase roll out of mandatory 2FA and concluded that UX was not impacted upon when using it for systems that dealt with sensitive information but the UX changed when its use was mandated for all systems. They focused on the UX experience especially on how to improve the interaction with 2FA but also made some recommendations related to trusted devices. Our University already implements conditional 2FA by white-listing devices as part of the roll out.

\section{Design Methodology and Limitations}
\label{sec:design}
\begin{figure}[t]
  \centering
  \includegraphics[width=0.8\linewidth]{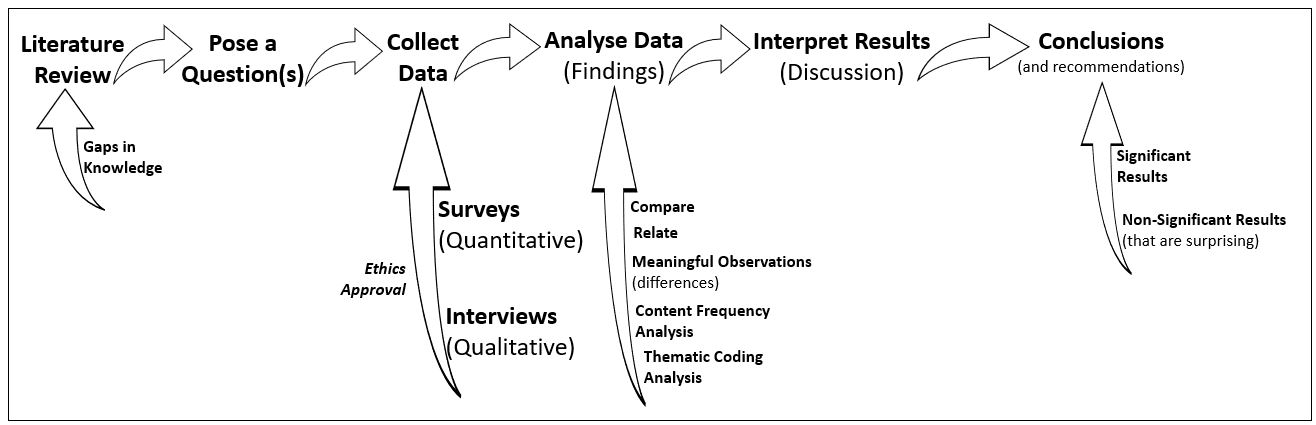}
  \caption{Mixed Method approach}
\label{fig:mixedmethod}
\end{figure}

In our research we adopted a mixed method approach~\cite{creswell:mixed}, as summarised in Figure~\ref{fig:mixedmethod}. We carried out an online survey with staff and students and semi-structured interviews with participants from the survey. We also carried out a number of interviews with the 2FA project IT team. We also analysed IT service request data.  

The staff and student input was gathered over a two month period from March 2020. The participants for our survey (n=273) and the subsequent interviews (n=11) were students and staff from the University. The number of participants that completed the survey was 226 and thus 47 participants were excluded from the final analysis due to failure to complete portions of the survey. The number of participants was low compared to the total University population, 
with 0.8\% (n=125) of students and 3.2\% (n=101) of staff. 
We recruited the online participants for the survey by using emails and intranet news bulletins. Those participants who were interviewed gave their details in the survey to permit us to follow up with them. 
The surveys were developed using the Qualtrics survey platform~\cite{qualtrics} and the semi-structured interviews were developed using an emergent approach based on the lessons learnt from the surveys. This allowed us to develop questions that could not be answered quantitatively having seen the quantitative results. It was important to adopt this emergent approach to avoid the challenge of when quantitative and qualitative results did not agree with each other.  We presented different questions in the interviews to the ones presented in the original survey to elicit more subjective perceptions and to gain further in-depth knowledge into the areas pertaining to our research questions. The individual user interviews were scheduled via email and were conducted online where the interviewer shared a screen and made interactive notes of a user's responses to the questions during the interview. The IT interviews were over a period of four months from June 2020 some in person and some online using a similar methodology.

All the processes related to the studies went through the University's ethical approval processes and participants were presented with guidelines on how we would process their data and they could withdraw at any point from the studies, preserving their anonymity.

We took insight and inspiration for the derivation of some of the questions of our survey from previous studies~\cite{DBLP:journals/corr/CristofaroDFN13,weidman:asac2017,Colnago:CMU}. There were 20 questions in total and none were mandatory and some open-ended questions were included. 
We did not conduct descriptive and inferential statistics due to the fact that the results may not be representative of the population and because we have categorical data. 
\subsection{Limitations}
The participants of the survey and interviews were self-selecting therefore we cannot make generalisations regarding the total university population. We only collected limited demographic data from the survey. We did not collect data on age, gender, or general computer usage. On reflection we should have done so to validate whether we had a diverse cross-section of respondents and in doing so we would also have been able to cross compare the confidence of using 2FA with the general IT expertise of the respondent and would then have been able to correlate IT expertise with perception of their value in terms of security and usability.

\section{Timeline for 2FA roll out at the University of Surrey}
\label{sec:timeline}
The University 2FA roll out is linked to Microsoft Azure Active Directory (AD) and applies to all systems that uses AD but does not cover legacy systems. It does cover Office 365 and the virtual learning environment which are considered to be key security attack points. The roll out was iteratively perform in phases as follows:

\begin{description}
\item[Phase 1]
 began with pilot users in August 2019. The pilot trials provided an opportunity to develop initial supporting documentation.  Initially, there was no default mandated for the second factor. The IT infrastructure supports the use of an authenticator app, SMS and hardware keys as a second factor in the authentication process. The users have the option of being able to use a Google Authenticator app but since the University uses Microsoft 365 the IT team recommend using the Microsoft Authenticator app to provide a more integrated user experience. The trials also served to identify the initial use cases that may cause friction during use for example when teaching, using lecture theatres and computer laboratories.
\item[Phase 2]
was the first large scale implementation and took place at the beginning of September 2019 for all students. The students were chosen as the first group of mandated users since it was easier to set this as an expectation on the student cohort at the beginning of an academic year. The key message during adoption was that 2FA was a way of improving the protection of systems from an individual’s perspective.
\item[Phase 3]
was the 2FA roll out for staff in Autumn 2019. Initially, the approach to staff deployment of 2FA was to allow an opt in policy when authenticating. Once the adoption figure slowed, this was changed to a mandatory use of 2FA to ensure full compliance. The mandated use by staff was lead with senior executives of the University being the first mandated adopters to reinforce the importance of adopting this new authentication technology and then in three groups based on the three faculties within the University. The same key message of  the protection of systems was highlighted as the driver for adoption. 
\item[Phase 4]
began in the Spring 2020 with the emergence of COVID and remote working being mandated for all staff and students. Hence,  a further iteration of the enforcement of mandatory 2FA was done but this was coupled with supporting everyone to be able to work remotely. It was during this phase that the use of an authenticator app was promoted as the default second factor. By July 2020 2FA was adopted by almost all of the 19,500 users within the University. There is a small group of staff for which there are exceptions,  i.e. users with no smart phones, some specialist Linux users and specific research areas. A small number of hardware-tokens have been provided as a second factor to ensure adoption of 2FA.
\item[Phase 5] 
(now post lockdown) is the continued deployment for staff and students in-year and its further mass uptake by all new students at the beginning of the new academic year in Autumn 2020.
\end{description}
Our user surveys and user interviews outlined in Section~\ref{sec:design} were conducted at the beginning of Phase 4 and our findings influenced the training updates made. The timing of our work was determined by the start of our research studies.

\section{Users' Findings}
\label{sec:findings}
Here we present results from the user survey and analysis from the user interviews conducted at the beginning of Phase 4.
\subsection{Training to use 2FA}
\label{usertraining}
When asked if respondents received formal training in the use of 2FA 83\% (n=187) responded to say they had not and those that did had been part of Phase 1. Only 22\% (n=49)  had received frequently asked questions (FAQs) information when asked to convert to 2FA and 65\% (n=148) did not receive any IT user support which meant that they had not logged any queries via the IT ticket support system to receive personal support.
The fact that respondents had not undertaken training has not impacted on the confidence of the majority of respondents in configuring an authenticator app with 73\% (n=165) indicating they were fairly confident or confident in setting up the authenticator app. Interviewees who were IT savvy noted that no instructions were needed, but others would have welcomed some formal training, as allowing people to figure it out themselves increased the level of confusion and stress. Moreover, some interviewees noted that the FAQs had missing steps/content and the training material was too long and overwhelming and not written in layman's terms. The general feedback included wanting a different mode of training through drop-in sessions, for example via MS Teams or interactively via animation to make the information more watchable and informative. Feedback from the interviews also included the need for more clarity on what to do when things go wrong in different scenarios, for example when someone loses their phone do they need to tell the university, what re-registration steps are needed, and what is coming next in terms of enhanced security technologies. 

\subsection{Usage of 2FA}
\label{sec:usage2fa:uni}
92\% (n=209) of respondents were using 2FA and this was down to the mandatory University requirement to use 2FA. Of the respondents 23\% (n=52) had been using it for 1-3 months but there was also 19\% (n=42) of experienced users that had been using it for a year in part due Phase 1 of the roll out. Notably the majority of the university respondents either used SMS, 71\% (n=160), or an authenticator app, 47\% (n=106) as the second factor. The feedback from the  interviews was that these two factors were easy to use and there was no awareness of other factors. There was strong willingness to use an authenticator app, with 63\% (n= 143) willing to use the university preferred authenticator app on their personal phone. However, 36\% (n=81) of respondents were generally unhappy about using their personal phone to download a work related app. Only 2\% (n=5) used tokens (e.g. Yubikey).

47\% (n=107) of respondents experienced usability issues with 2FA but a comparable number, 43\% (n=98), of respondents did not. Of those experiencing issues it was largely due to an authenticator app not working, not receiving SMS messages due to poor connectivity and changing devices. 
34\% (n=76) of respondents found it easy to use whereas 10\% (n=22) found it difficult to use but  57\% (n=128) found it annoying to use. Respondents and interviewees commented on the heavy reliance on the phone and not being able to predict when you would need to re-authenticate.

When asked whether respondents used 2FA to log into systems other than university systems, 60\% (n=136) said yes but 33\% (n=75) said no. This suggests that there is wider adoption of 2FA and may explain why 53\% (n=119) would recommend the use of 2FA and only 23\% (n=51) would not. 
The most popular second factors for non-university systems is the same as university systems, with 47\% (n=106) of respondents preferring SMS and 36\% (n=82) of university respondents preferring the authenticator app. 8\% (n=17) of respondents used hardware tokens for non-university systems and thus is also the least popular second factor.
The feedback from interviewees was that there was little awareness of hardware keys or of how to use them. For those that had an awareness, there were concerns about having to carry them on their person, the potential of losing them and the expense of having to replace them. Of those using hardware keys they state the inconvenience that not all browsers supported them, for example Safari and Microsoft's older browsers.

\subsection{Security Perceptions of 2FA}
\label{sec:security2fa:uni}
80\% (n=181) understood why 2FA was rolled out at the University and that it was to improve the security posture of the organisation. 70\% (n=158) of respondents indicated that there is no change in their perception of the need to use 2FA. Similarly, the perception of the benefit of using 2FA has not changed for 68\% (n=153) of the respondents following the roll out.  

Several of the interviewees highlighted the security threats with SMS but that it was easy to use. 82\% (n=185) of university respondents felt that their accounts are less likely to be compromised having enabled 2FA. Some interviewees when asked to comment about the findings expected that adding an extra layer of security would make people more comfortable and that the figure should be higher whereas 28\% (n=64) of university respondents felt they did not have to worry as much about account security having enabled 2FA. 

67\% (n=152) respondents used the ``remember me'' option on their devices for 14 days when logging in because it was convenient. Thus, this means that the computer (or the browser) plays the role of the authenticator app. The majority of the interviewees recognised that this meant that authentication attempts from a remembered computer could be done without a second factor; this improved the user experience but at the trade-off of security. Some of the interviewees, however, did not want the browser to store their personal credentials and would rather enter their details at every login. But the annoyance of having to regularly response to a 2FA challenge or being restricted to having the remember me functionality for only 14 days was highlighted in the interviews as affecting usability. 
 
About two-thirds, 63\% (n=142) of people said that within the context of 2FA they would value ‘security’ the most, while only a quarter, 24\% (n=55) valued ‘usability’ the most, and less than 3\% (n=7) said ‘trust’ and ‘privacy’.  There was no consensus from the  interviewees on what was meant by trust and privacy in the context of 2FA.

\section{Lessons Learnt from the 2FA roll out from an IT perspective}
\label{sec:ITfindings}
\subsection{Training}
The IT team held weekly project meetings and took an agile approach to fixing errors and updating the training material depending on the resources available. Most of the improvements were on making things clearer in the user documentation online and providing individual support to those that needed it.

In Phase 3 the senior executives of the University received personal training and were guided to the user documentation and the benefits of 2FA were explained to them. However, they performed the set up themselves which was a test of the usability of the user documentation but the lesson learnt was that the benefit of using 2FA needed to be clearer and made more prominent in the communication to those adopting the 2FA security mechanism. Thus, the documentation and set up information was amended to make this clearer.

 In Phase 4 it was clear that once the COVID situation emerged and staff and students were required to work remotely there needed to be a significant improvement of the user documentation on the intranet. Training videos were created to step through the set up processes in a visual way. Moreover, clearer FAQs were added  for example to address the loss or change of device and the loss of an SMS signal. The user documentation also included how to use 2FA in conjunction with wanting to use the virtual private network (VPN) to connect to the work systems remotely. Therefore, the insight was that it was important to include documentation regarding the context within which they would be using the security mechanism.

In Phase 4 there was continued reflection on the use cases of where people were using their devices. The user documentation was improved to reflect that fact that the second factor settings can be changed. For example, if a user would be travelling and would not have access to an SMS signal the settings can be configured to take this into account. The aim of this was to improve users awareness of how to use 2FA when travelling or when their remote working location would require different second factor settings. 

Since the roll out was phased it meant that the authenticator app was not initially the default second factor and there are many users who use SMS as their preferred second factor. It is not possible to track who is using the authenticator app or SMS and therefore it is difficult to get users to move over to using the app as their primary second factor. Therefore, in Phase 5 new communication is needed to improve awareness of users that the recommended default second factor is the authentiactor app but it is important to do this in a timely and visible way. This is being done as part of cyber awareness month within the University in Autumn 2020 so that it is coupled with raising awareness of good security practices otherwise it could be a recommendation that simply gets lost in the midst of other emails. 
\subsection{User support}
The most significant amount of user support required was at the beginning of Phase 1 when students started adopting 2FA and in Phase 4 when users were initially working remotely.

In Phase 2 a lesson learnt from student adoption was when students were registering for 2FA overseas and then when they came to the University and used different devices this meant they would have to go through the process of re-registering. Hence it provided an initial poor experience when they arrived on campus in the UK. Thus, in the user documentation and set up processes guidance on remote set up was quickly amended and was highlighted to both students and staff in the communication when setting up 2FA for the first time.

In Phase 4 there were three case requiring individual support: one of an elderly colleague who had an old mobile phone and two that did not wish to use their own phones. The IT team commented on the need to be sensitive to users who may not have access to the newest phones nor the desire to the use them and to work patiently with those users to gain a picture of what they were doing remotely in order to get them set up. It was also important to be able to provide provision to those that did not want to user their own phone so that they were not discriminated from being able to use the University systems. Hence, the University provided hardware tokens for around 10 users to use.

The IT Team also understood how reading out approval requests and reading out the details from a hardware token for the visually impaired would work. However, there have been no specific IT requests from users with disabilities but they had planned for this. This was an example of good planning from a user experience perspective. 

\subsection{Usability of 2FA}
In Phase 1 the IT team identified a use case were there was a need to provide conditional 2FA access if students/staff were on managed devices on campus. These devices are ones that are in lecture theatres and laboratories managed by the IT team. For example, if students were in a laboratory then it was important to provide a seamless experience. Similarly when staff were logging into a computer in a lecture theatre the user experience needed to be frictionless and they should not be prompted for additional authentication steps during teaching. This means those devices needed to be trusted to exclude the 2FA requirement. Some of the initial teething problems with conditional 2FA were related to users experiencing a 2FA challenge whilst authenticating on a trusted device. This was linked back to the trusted platform modules (TPM) not being turned on in the BIOS of the computers in the lecture theatres and laboratories resulting in the computer not completing a hybrid join to Azure Active Directory. A lesson learnt from this was the need to conduct an audit of the managed estate prior to roll out. 

In the remainder of this section we focus on reflections of our IT team of some of our survey findings.  We indicated that several of the early adopters chose to use SMS as the second factor after they had trouble downloading and using the authenticator app. In our discussions with the IT team no one was aware of users raising such issues through IT service requests.  When we discussed what guidance there should be for a user who could not get access to SMS the IT team's advice was to revert to the authenticator app since this did not require a data connection. The IT team felt that there was a balance between keeping the instructions simple for the majority of users and covering all eventualities in the user documentation. It was nonetheless recognised that there would be special cases and that it was very important to provide individual user support for these users.

From the quantitative results in Section~\ref{sec:findings} we noted that  47\% were annoyed with using 2FA but the feedback from the IT team was that only a few people raised queries with the deployment team and that they were all handled successfully by explaining the reasons of why it was needed.  We also noted that 36\% of people were unhappy about using their own phones but again the IT team commented that only very few (2-3) people had raised this through an IT service request. It was felt that due to COVID people were more willing to use their own devices because they had to just get on with getting connected in order to work. Again the IT team did provide individual support where needed and for example they have issued around 10 hardware tokens during lockdown to users (phase 4) and one post lockdown (phase 5) to those that strongly did not wish to use their own phones. 

Users also requested that re-authentication period be longer than 14 days. The IT team commented that this was the recommended duration by Microsoft. As part of reflecting on when authentication approvals were being made a more risk based approach was taken that was sensitive to the location of where a user logged in. If a user logged in from an unknown location they would be prompted to re-authenticate using 2FA. The mechanism for these additional authentication challenges was rolled out in June 2020 to further improve the security posture of the organisation when users were working remotely.

The University has an accessibility and audit programme for reviewing web pages but reviewing the processes for a good 2FA user experience for students and staff with disabilities was not initially covered as part of the roll out. The University is now reviewing the set up pages for 2FA as part of an audit programme based on accessibility and assisted digital guidelines (WCAG 2.1)~\cite{wag}. The insight gained by the IT team is to continually assess whether an IT deployment addresses the wider user experience needs of all users so that for each iteration of an IT deployment there is an opportunity for improving the user experience.

\subsection{Impact on IT team}
The above has provided reflections on how users were supported by the IT team but there are also important lessons learnt for the IT team as they gained experiences of a deployment that was very user centric. The initial phases of the roll out were focused on getting the functionality working and did not focus on how the security mechanism could be misused, what weaknesses needed to be overcome and how to get the maximum benefit of the phased roll out. Before the roll out the IT team first needed to evaluate whether there were weaknesses in the wider IT infrastructure before a deployment could take place and only then could the initial functionality be tested. In the initial phases (1-3) only standard users adopted 2FA, i.e., not users that would challenge the set up, for example Linux users or those that worked remotely in areas of poor signal or travelled a lot. Therefore, an important lesson learnt was the need to include in the pilot and initial phases a cross section of  users that used different devices and operating systems or who worked remotely, including overseas, to fully understand how users interacted with the security mechanism. The other lesson learnt was the impact on a deployment given a wider IT context. In Phase 4, as a consequence of a large number of users working remotely due to the COVID situation, the University changed its VPN platform to improve its security posture. Therefore, ensuring that 2FA worked with a new VPN could not have been planned for initially. Thus, the IT team quickly changed their working style to collectively solve problems but this required colleagues to adapt their working styles and methods of communication. 

\subsection{Business Impact and Metrics}
The IT team are primarily monitoring user experience  post-deployment of 2FA through the number IT service requests that are being made and the issues that are raised. The number of IT service requests related to authentication was 10.4\% (n=856) in Q4 of 2019 of the total number of IT service requests, i.e., when the majority of students were mandated to use 2FA. Then number of calls related to authentication has continued to drop from 7\% (n=580) in Q1 and 5.6\% (n=279) in Q2 when staff were mandated to use 2FA, and 4.1\% (n= 280) in Q3 of 2020. The majority of all requests were related to 2FA reset requests which provides insights on the importance on the clarity of user documentation for this aspect. 

Notably, the IT team has recorded that the number of service requests related to account breaches linked to phishing attacks logged have reduced by 76\% by the end of Phase 3 compared to prior to Phase 1. This is an important commercial driver in academia and is a direct cost benefit of a 2FA roll out.

More generally, there are two further interesting insights gained. Firstly, it was no longer be possible for the personal assistants of some senior executives to be able to log into executives' emails to provide support. Hence, alternative working practices needed to be put in place that maintained the security of individuals' accounts. 
Secondly, the IT team faced the challenge of needing to be familiar with a complete range of both corporate and personal devices and phones because users use a plethora of devices. The challenge was not the scale of the roll out nor the fact there were different kinds of users of users but the sheer number of devices that the IT team needed to be aware of in the cases of user queries. This challenge may be broader than in corporations where they can be more prescriptive in what they allow.  

\section{Discussion}
\label{sec:discussion}
We recognise that when using an authenticator app as a second factor the wider phone ecosystem needs to be considered in order to evaluate perceived usability, as did Weidman and Grossklags~\cite{weidman:asac2017}. The key significant finding from the survey was that SMS is the predominant second factor and the use of hardware keys is the least popular.  For those using SMS there were concerns of not being able to use SMS in areas of poor connectivity. The significant use of SMS as a second factor was a surprise to the IT team and this led them to change the default option to making the authenticator app the default option whenever possible in Phase 4 and in future roll outs.  No detailed investigation has been conducted as to the nature of why people did not get the authenticator app working initially nor whether a backup authenticator app was set up as an alternative to those using SMS. 

The roll out has gone very well and the iterative approach to improve the training and user documentation during the roll out was successful. One point of improvement is the fact that there is  no way to monitor which second factor people had chosen to use initially since they were given flexibility and therefore it is difficult to ensure that everyone is now using the preferred second factor of the authentication app. One way to address this is to produce new communication with all users ask them to confirm whether they have configured 2FA properly so that those who are using SMS can be identified. 

The concerns raised by users in our findings on what to do when they lost a device or changed to a new device has been clearly addressed in updates to the user documentation in Phase 4 as well as the method of training included videos as well as FAQs.

As we have mentioned earlier the annoyance of having to regularly respond to a 2FA challenge or being restricted to having the remember me functionality for only 14 days was highlighted by users. The inherent security risks with the remember me process have been recently highlighted by Patat and Sabt~\cite{rememberme}. Thus, the wishes of interviewees to be able to customise the duration of the remember me and not having to respond to 2FA challenges even if browser cookies are cleared would not make sense to deploy in the context of associated security risks. Thus this is not something that can be solved by training. But with increased cyber awareness of the wider security context of using 2FA users could learn to accept this as being a necessary restriction. Therefore, the open point of discussion is how much context to put in the user documentation to ameliorate these concerns and how such issues should be addressed by a wider cyber awareness training programme. The majority of the respondents recognised that the use of 2FA would improve an organisation's security posture and it was interesting that the university staff and students valued security more than usability. 

\section{Conclusion}
In our case study we identified several key lessons including the fact that a pilot roll out prior to mass deployment enabled the IT team to come up with many what if scenarios and to plan for them but that it was important for a pilot to include many different kinds of users. Moreover, by adopting an agile approach it was possible to iteratively update user documentation as new use cases emerged on the kind of support information users needed, and these updates were done by the IT and website team within the organisation. Also the training material needed to be a mixture of website material and step by step media guides. It was also key that the users who needed personal support could be identified and supported individually so that they could adopt the security mechanism to avoid the potential for digital exclusion. Interestingly the user evaluation highlighted how users felt but that this did not translate into IT service requests. Therefore, is was important to carry out the user evaluation so that it could influence user documentation to support improving users' perception of the benefits of the technology and why some of the usability restrictions were in place. Our user evaluation also highlighted the importance of training users not only in how to use the security mechanism being deployed but to explain why it was needed in the wider context of good security practices.

\label{sec:conc}

\noindent 
\section*{Acknowledgments}
Abideen Tetlay is supported by the UK Government PhD studentship scheme. Thanks to the participants, the IT team and the input from Sociology colleagues at the University on our methodology.




\bibliographystyle{splncs04}
\bibliography{bibliography}


\end{document}